\documentclass[twocolumn,prb,showpacs,superscriptaddress,showkeys,preprintnumbers,amsmath,amssymb]{revtex4}

\usepackage{graphicx}
\usepackage{dcolumn}
\usepackage{bm}
\usepackage{multirow}
\usepackage{subscript}
\usepackage{nicefrac}

\begin{document}


\title{Crystal and magnetic structure of the oxypnictide superconductor LaO$_{1-x}$F$_x$FeAs: evidence for magnetoelastic coupling}

\author{N. Qureshi}

\email[Corresponding author. Electronic
address:~]{qureshi@ph2.uni-koeln.de} \affiliation{$II$.
Physikalisches Institut, Universit\"{a}t zu K\"{o}ln,
Z\"{u}lpicher Strasse 77, D-50937 K\"{o}ln, Germany}

\author{Y. Drees}
\affiliation{$II$. Physikalisches Institut, Universit\"{a}t zu
K\"{o}ln, Z\"{u}lpicher Strasse 77, D-50937 K\"{o}ln, Germany}

\author{J. Werner}

\affiliation{Leibniz-Institut f\"{u}r Festk\"{o}rper- und
Werkstoffforschung (IFW) Dresden, D-01171 Dresden, Germany}

\author{S. Wurmehl}

\affiliation{Leibniz-Institut f\"{u}r Festk\"{o}rper- und
Werkstoffforschung (IFW) Dresden, D-01171 Dresden, Germany}

\author{C. Hess}

\affiliation{Leibniz-Institut f\"{u}r Festk\"{o}rper- und
Werkstoffforschung (IFW) Dresden, D-01171 Dresden, Germany}

\author{R. Klingeler}

\affiliation{Leibniz-Institut f\"{u}r Festk\"{o}rper- und
Werkstoffforschung (IFW) Dresden, D-01171 Dresden, Germany}

\author{B. B\"{u}chner}

\affiliation{Leibniz-Institut f\"{u}r Festk\"{o}rper- und
Werkstoffforschung (IFW) Dresden, D-01171 Dresden, Germany}

\author{M. T. Fern\'{a}ndez-D\'{i}az}

\affiliation{Institut Max von Laue-Paul Langevin, 6 rue Jules
Horowitz, BP 156, 38042 Grenoble Cedex 9, France}

\author{M. Braden}

\email{braden@ph2.uni-koeln.de}

\affiliation{$II$. Physikalisches Institut, Universit\"{a}t zu
K\"{o}ln, Z\"{u}lpicher Strasse 77, D-50937 K\"{o}ln, Germany}

\date{\today}

\begin{abstract}

High-resolution and high-flux neutron as well as X-ray
powder-diffraction experiments were performed on the oxypnictide
series LaO$_{1-x}$F$_x$FeAs with $0\le x\le0.15$ in order to study
the crystal and magnetic structure. The magnetic symmetry of the
undoped compound corresponds to those reported for ReOFeAs (with
Re a rare earth) and for AFe$_2$As$_2$ (A=Ba, Sr) materials. We
find an ordered magnetic moment of 0.63(1)~$\mu$\textsubscript{B}
at 2 K in LaOFeAs, which is significantly larger than the values
previously reported for this compound. A sizable ordered magnetic
moment is observed up to a F-doping of 4.5\% whereas there is no
magnetic order for a sample with a F concentration of $x$=0.06. In
the undoped sample, several interatomic distances and FeAs$_4$
tetrahedra angles exhibit pronounced anomalies connected with the
broad structural transition and with the onset of magnetism
supporting the idea of strong magneto-elastic coupling in this
material.

\end{abstract}


\maketitle

\section{Introduction}
\label{sec:Introduction}

\begin{figure}
\includegraphics[width=0.5\textwidth]{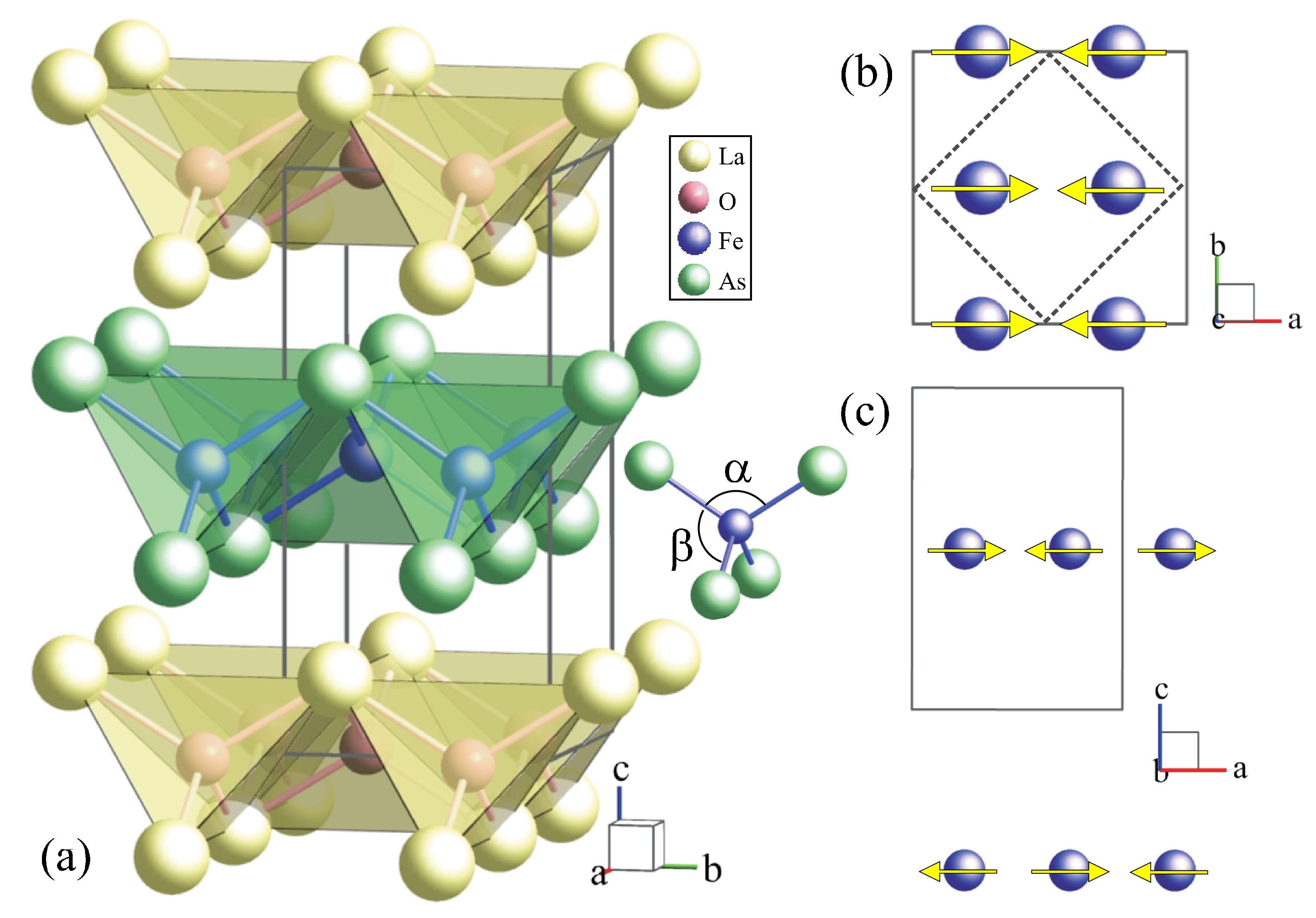}
\caption{\label{fig:structure} (Color online) (a) Visualization of
the tetragonal crystal structure of LaO$_{1-x}$F$_x$FeAs and the
definition of the As-Fe-As bond angles. (b) The magnetic structure
only showing the Fe ions in the orthorhombic unit cell (straight
line). The dashed line depicts the tetragonal cell. (c)
Alternation of the moment direction along the {\it c} axis due to
the propagation vector $\mathbf k$=(1 0 $\frac{1}{2}$)}
\end{figure}

The recently discovered family of oxypnictides
superconductors\cite{kam2008} has focused the interest of the
scientific community as they represent the first non-copper-oxide
based layered superconductors reaching a $T_c$ of 55
K.\cite{ren2008} Their crystal structure is similar to the one
adopted by the copper-based superconductors i.e. a layered
structure where FeAs sheets are sandwiched by LaO/F sheets
(Fig.~\ref{fig:structure}(a)).
Like for the cuprates superconductivity arises by chemical doping
and suppression of the magnetic ground state; however, also the
application of pressure to the non-doped system can induce
superconductivity for certain FeAs
compounds\cite{tor2008,ni2008,ali2009,par2008,kre2008} in clear
contrast to the cuprates where the antiferromagnetic state of the
parent phase is a Mott-Hubbard insulator requiring electronic
doping in order to obtain metallicity and superconductivity. For
SmO$_{1-x}$F$_x$FeAs the magnetic ordered state even extends to
doping levels within the superconducting regime and low-energy
spin fluctuations have been observed up to the doping levels where
$T_c$ is maximal.\cite{dre2009} These findings suggest an
important role of magnetism in the superconducting pairing.

The magnetism in the FeAs compounds appears to be very sensitive
to the structural details which in turn modify the Fermi nesting
conditions and the geometric frustration. It has even been stated
that the structural distortions play a more important role in the
modification of the Fermi surface than charge doping for inducing
superconductivity.\cite{kim2009} The shape of the FeAs$_4$
tetrahedra seems to be decisive, as the highest superconducting
transition temperatures are obtained for regular FeAs$_4$
tetrahedra.\cite{lee2008,zha2008} This observation is corroborated
by density functional theory calculations which reveal a clear
dependence of the Fe magnetic moments as well as of the magnetic
interaction parameters on the shape of the FeAs$_4$ layers, i.e.
the FeAs bond distance and the layer thickness.\cite{yin2008}

We have combined high-flux and high-resolution neutron and X-ray
powder diffraction experiments to study the magnetic and crystal
structure of the LaO$_{1-x}$F$_x$FeAs series. We may unambiguously
determine the magnetic symmetry of the undoped material finding a
sizeable ordered moment. The doping dependence of structural
parameters qualitatively confirms earlier studies, but upon
cooling  through the structural and magnetic transitions, the pure
and slightly doped materials exhibit anomalies in bond distances
and in bond angles which reflect the general magnetophonon
coupling in FeAs compounds.


\section{Experimental}
\label{sec:Experimental}

Powder samples of LaO$_{1-x}$F$_x$FeAs have been synthesized using
a previously reported two-step solid-state reaction
method.\cite{kam2008,zhu2008,kon2009} We mention that for our
sample with the nominal F content $x$=0.04 investigated in this
work, wavelength dispersive X-ray analysis yields a higher F
content as compared to a sample with the same nominal
concentration which has been investigated in previous studies of
our group, see for example \cite{kon2009,lue2009,wang2009}. We
therefore refer to the former sample by using the measured F
content ($x$=0.045) in order to discriminate it from the latter
and otherwise use the nominal doping levels throughout the paper.

Moreover, we note that for our sample with x=0, resistivity
measurements suggest \cite{kon2009,Hess2009} $T_S\sim152$~K (peak
of $\rho(T)$) and $T_N\sim 135$~K (inflection point of $\rho(T)$),
i.e., at somewhat lower temperatures as have been found in our
previous undoped sample \cite{kon2009,lue2009,Hess2009,wang2009}.

The nuclear structure investigation of all powder samples
(0$\leq$x$\leq$0.15) at room temperature (RT), 200 K, 100 K and 2
K has been carried out at the high-resolution neutron powder
diffractometer D2B (Institut Laue-Langevin, Grenoble), where the
$x$=0 and 0.045 samples have been examined at additional
temperatures. The experiments have been performed using the
wavelength of 1.594~\AA{} from the Ge(335) reflection. Powder
diffraction patterns have been recorded with a counting time of 3h
per temperature point. The magnetic structures of the samples with
$0\le x\le0.06$ have been investigated at the high-flux neutron
powder diffractometer D20 (ILL). A wavelength of 2.41~\AA{}
emerging from the (002) reflection of a pyrolytic graphite
monochromator has been used, which gives a good compromise between
high flux and reciprocal space resolution. Diffraction patterns
have been recorded for 2h above the magnetic transition and at
various temperatures within the magnetically ordered phase. The
same samples have been used for further experiments at an X-ray
powder diffractometer Siemens D5000. Each X-ray diffraction
pattern has been measured for 5h using CuK$\alpha$ radiation.

\section{Results and Discussion}

\subsection{ Magnetic structure in pure and slightly doped  LaOFeAs }

The magnetic structure of LaOFeAs has not been unambiguously
determined so far. First evidence for a spin-density wave
instability was obtained from kinks in the resistivity and in the
magnetic susceptibility further supported by electronic-structure
calculations.\cite{dong2008} The diffraction experiments by De la
Cruz et al. showed that a structural phase transition occurs
slightly above the magnetic transition. The structural and
magnetic transition temperatures amount to T$_S$=155\ K and
T$_N$=137\ K respectively.\cite{cru2008} However, the nuclear and
magnetic structures proposed in Ref.~\onlinecite{cru2008} appear
very unlikely. As justified in detail below and in agreement with
the findings for ReOFeAs and AFe$_2$As$_2$ we will assume an
orthorhombic and not a monoclinic low-temperature phase in the
following. The orthorhombic $a_o$ and $b_o$ axes are rotated by
45$^o$ degrees with respect to the tetragonal axes $a_{tet}$. The
proposed magnetic structure consists of ferromagnetic stripes of
neighboring Fe moments running along an orthorhombic axis
antiferromagnetically stacked along the perpendicular axis, see
Fig.~\ref{fig:structure}(c). In the model proposed in
Ref.~\onlinecite{cru2008} the stacking vector $q_{stack}$ is
perpendicular to the magnetic moment $m_{Fe}$ which amounts to
0.36(5)~$\mu$\textsubscript{B} at T=8\ K. A following neutron
diffraction experiment on LaOFeAs by McGuire et al. \cite{mcg2008}
also reports a moment of 0.35~$\mu$\textsubscript{B} but does not
discuss the spin orientation. Huang et al. describe a magnetic
structure with $q_{stack}$ parallel to the magnetic moment
$m_{Fe}$, which disagrees with the earlier report,\cite{cru2008}
but apparently the instrument resolution is insufficient to
determine the spin direction along $a_o$ or $b_o$.

\begin{figure}
\includegraphics[width=0.45\textwidth]{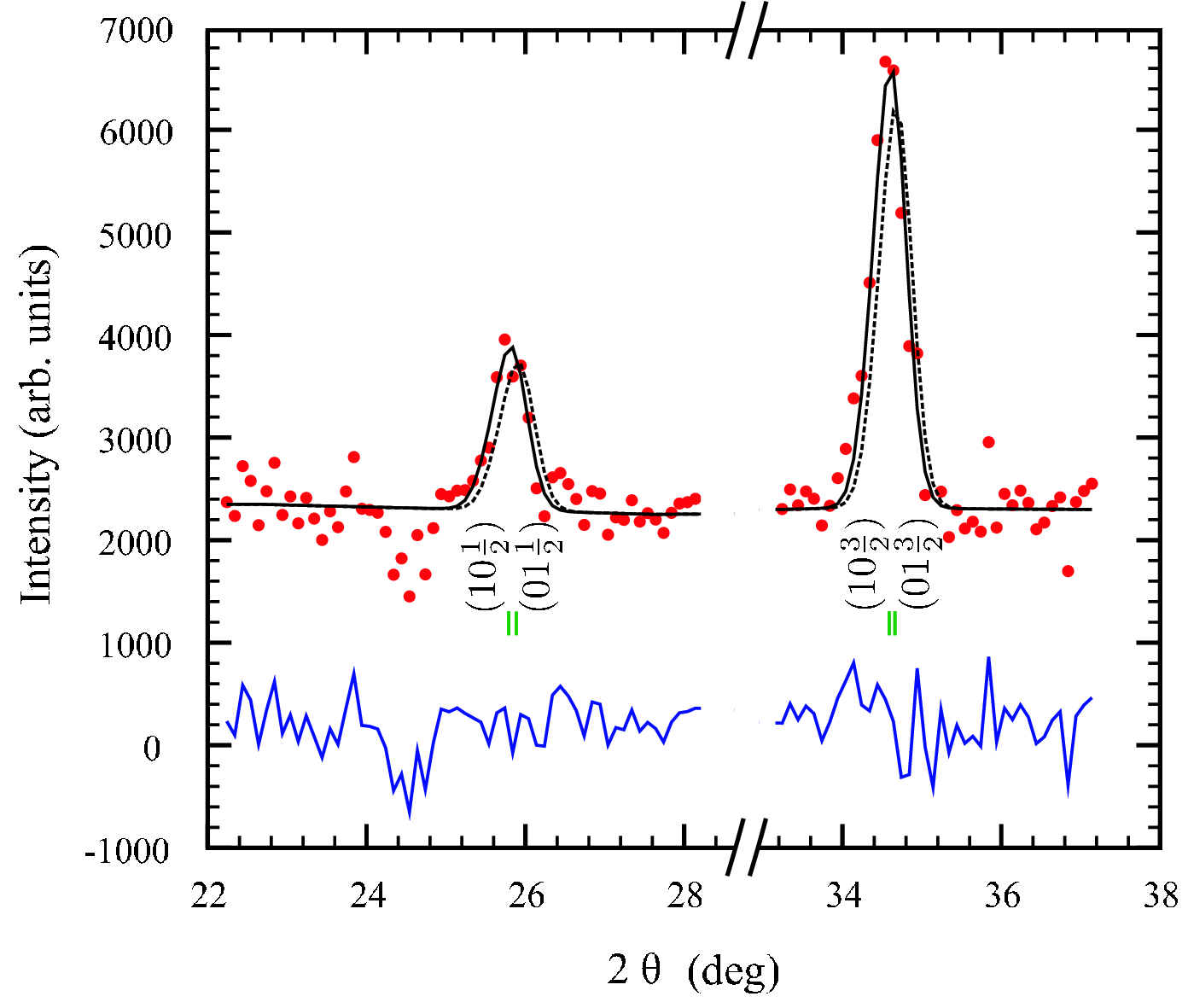}
\caption{\label{fig:mag} (Color online) Extract of the observed
pattern (dots), calculated patterns (upper lines) and difference
plot (lower line) of two magnetic models for LaOFeAs. The
calculated patterns result from magnetic structure models with the
Fe magnetic moments along the {\it a} axis (straight line) and
along the {\it b} axis (dashed line)}
\end{figure}

We have analyzed all powder diffraction patterns using the
FullProf program.\cite{fullprof} The structural investigation at
the high-resolution diffractometer D2B confirmed the correct phase
formation of the tetragonal $P4/nmm$ and orthorhombic $Cmme$
structures above and below the structural phase transition,
respectively, see below. In order to focus on the weak magnetic
scattering the D20 diffraction patterns of the paramagnetic phases
have been subtracted from the respective patterns of the
magnetically ordered phases. For the undoped sample the
differences are manifest in two well defined magnetic peaks,
indexed as $(1 0 \frac{1}{2})$ and $(1 0 \frac{3}{2})$ at
2$\theta=25.8^\circ$ and 34.6$^\circ$, respectively, to which a
magnetic structure model has been refined (Fig.~\ref{fig:mag}).
Representation analysis\cite{chatt,group3} has been used to derive
those magnetic structure models which are compatible with the
nuclear structure ({\it Cmme}) and with the propagation vector
with a half-indexed $c^*$-component and an antiferromagnetic
coupling between two Fe moments related by the C centering: (1 0
$\frac{1}{2}$).\cite{note} Out of the six possible collinear spin
configurations (three directions along the principal
crystallographic axes and the coupling between the two moments
chosen at (0.25,0.0,0.5) and (0.75,0.0,0.5) in orthorhombic
notation) only two may describe the data sufficiently well. These
are the models with $q_{stack}$ parallel to the magnetic moment
$m_{Fe}$. Since neutron diffraction only measures the magnetic
component perpendicular to the scattering vector, the models with
$q_{stack}$ perpendicular to the magnetic moment $m_{Fe}$ proposed
in Ref.~\onlinecite{cru2008}, can be clearly excluded. To decide
between the alignment of the magnetic moments along $a_o$ or $b_o$
is more difficult, but due to the high statistics and resolution
of the D20 data this is possible as well.
The  solid (black) and dashed lines in Fig.~\ref{fig:mag} show the
calculated patterns corresponding to the models where the moments
are aligned along the $a_o$ axis and along the $b_o$ axis,
respectively. As the  $b_o$ axis is 0.4\% shorter compared to
$a_o$ the two models can be distinguished.  The model with the
magnetic moments along the {\it a} axis is better suited to
describe the observed peaks which is also expressed by the
respective R values (10.71\% for $\mu\vert\vert${\it a} and
12.76\% for $\mu\vert\vert${\it b}). The Fe magnetic moment along
the {\it a} axis has been refined to
0.63(1)~$\mu$\textsubscript{B} which is about twice the size of
the magnetic moment presented in Refs.~\onlinecite{cru2008}
and~\onlinecite{mcg2008}. The ordered moment at the Fe site still
seems to be substantially lower than in other undoped FeAs
compounds \cite{lynn-dai2008} but the difference is much smaller
than previously reported.

\begin{figure}
\includegraphics[width=0.4\textwidth]{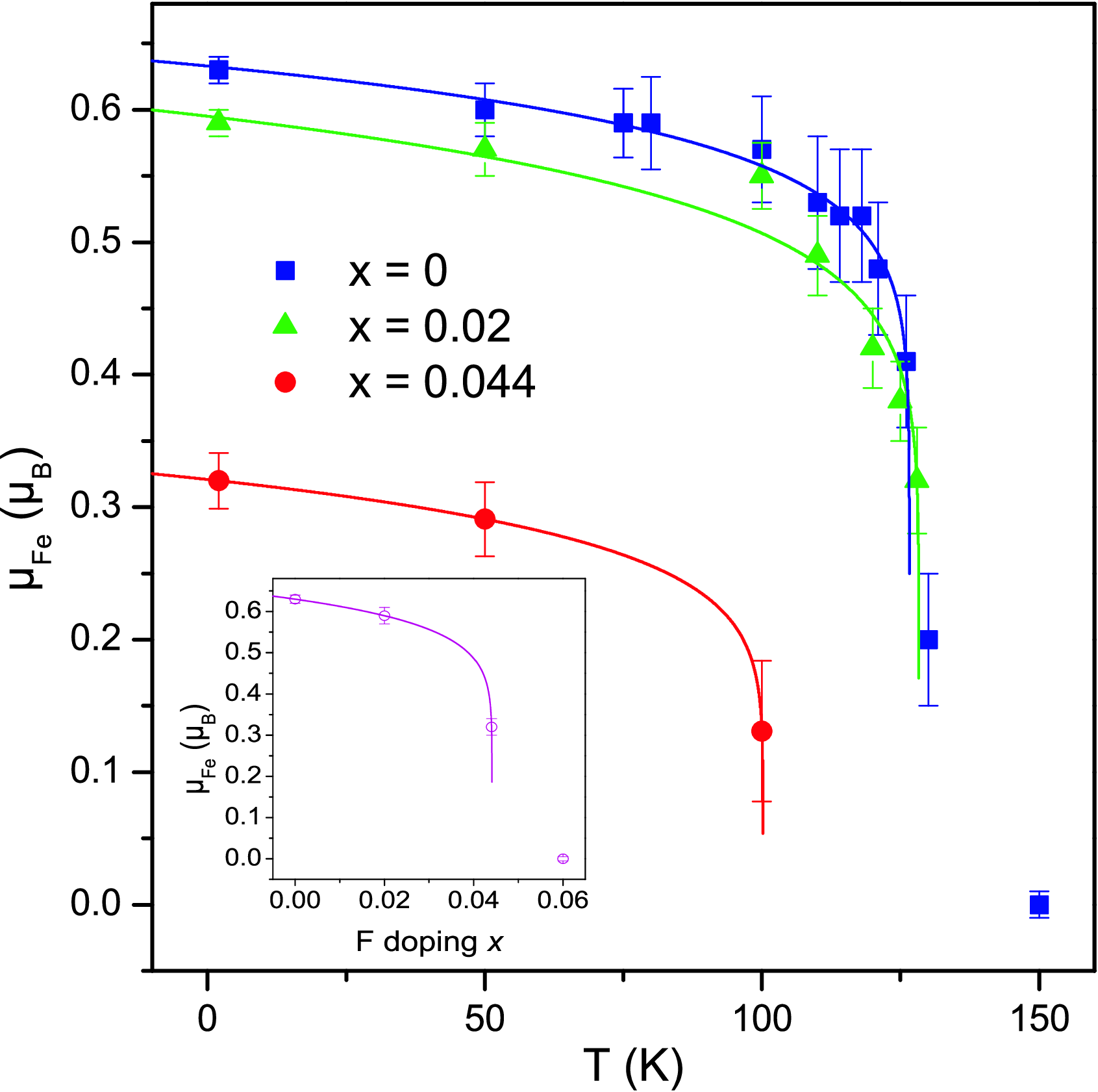}
\caption{\label{fig:moment} (Color online) Magnetic moment of the
Fe ions in dependence of temperature and F doping (inset, 2 K).
Power law functions have been fitted to the data as guides to the
eye (solid lines).}
\end{figure}

\begin{figure}
\includegraphics[width=0.45\textwidth]{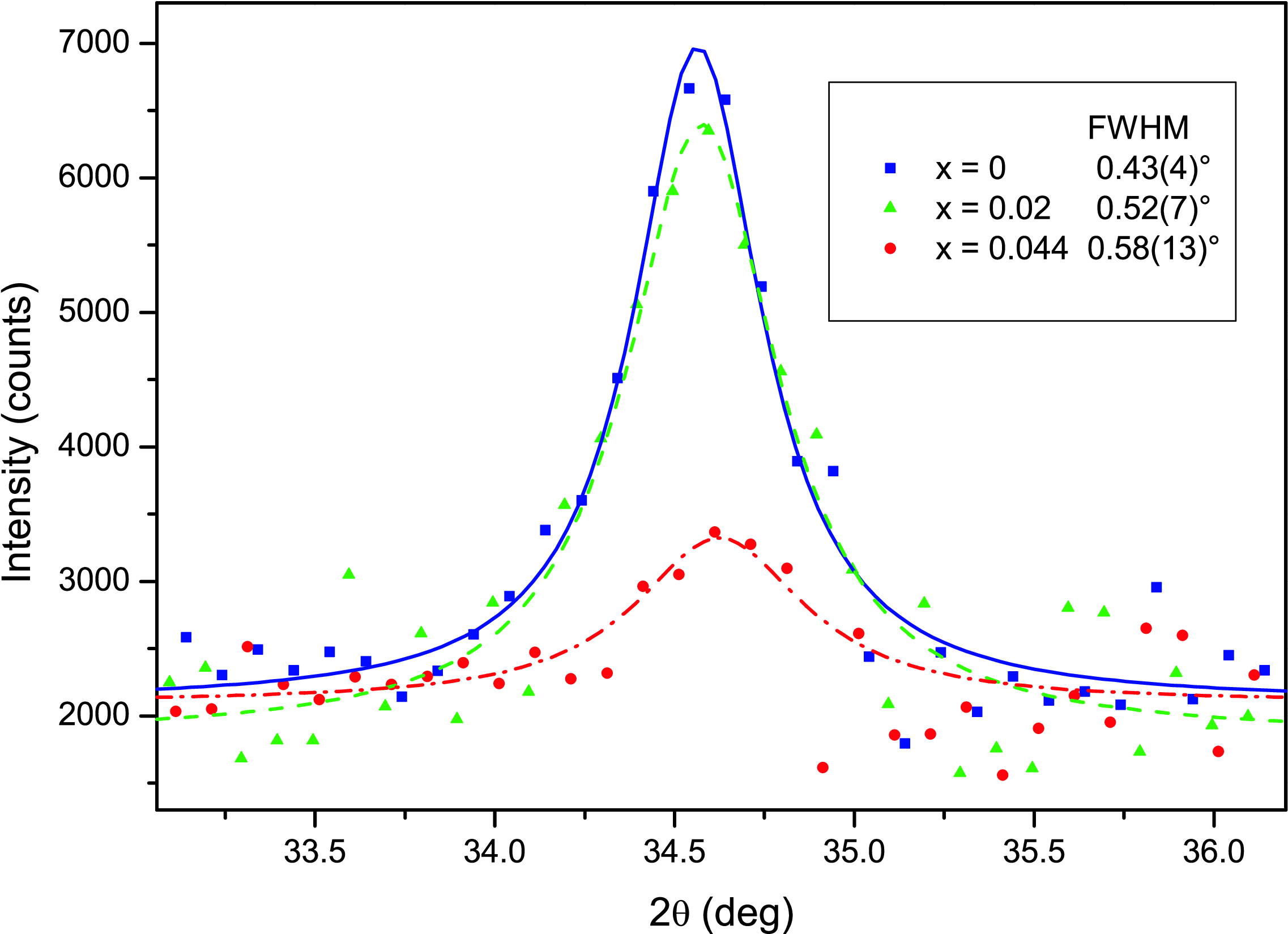}
\caption{\label{fig:fwhm} (Color online) Data points show the
magnetic (10$\frac{3}{2}$) reflection of the $x$=0, 0.02 and 0.045
samples. The FWHM have been extracted from Pseudo-Voigt fits and
reveal a broadening with increasing doping level.}
\end{figure}

The same magnetic model was used to describe the data for the
slightly doped samples. The refined magnetic moments for the
$x=0.02$ and $x=0.045$ samples are 0.59(2)~$\mu$\textsubscript{B}
and 0.32(2)~$\mu$\textsubscript{B}, respectively. For samples with
$x\geq0.06$ no magnetic scattering could be observed. The three
powder samples revealing magnetic order have been investigated in
more detail as a function of temperature. Fig.~\ref{fig:moment}
shows the temperature and doping dependent suppression of the
magnetic order.
One can observe that the undoped and the $x$=0.02 sample exhibits
similar magnetic moments with a similar temperature dependence,
but the magnetic reflections of the doped sample exhibit a
slightly broader full width at half maximum (FWHM) than those of
the undoped sample (Fig.~\ref{fig:fwhm}). A decrease of the
coherence length would be in perfect agreement with the subsequent
suppression of magnetic order, which is then expressed by the
strongly reduced magnetic moment for $x$=0.045 and finally by the
absence of long-range order for $x$=0.06. However, we may not
fully rule out that the broadening of the magnetic peaks is due to
an incommensurate magnetic order; but in case of the $x$=0.02
sample such an effect must be small.

\subsection{ Crystal structure in LaOFeAs }

The crystal structure of LaOFeAs was studied by high-resolution
neutron diffraction and by x-ray powder diffraction. The
structural phase transition forms the most prominent feature in
the temperature dependence of the crystal structure. In agreement
with most previous studies we find that the low-temperature phase
is orthorhombic, space group $Cmme$; the refinement of the
structure model in the monoclinic space group proposed in
Ref.~\onlinecite{cru2008} did not yield a better data description;
the structural parameters given in Ref.~\onlinecite{cru2008}
clearly worsen the reliability factors. However, the phase
transition from $P4/nmm$ to $Cmme$ seems to be quite uncommon.
This transition is of the proper ferroelastic character and it can
be directly related to the magnetism. The magnetic order at low
temperature exhibits an orthorhombic symmetry due to the alignment
and the stacking of the ferromagnetic stripes and due to the fixed
orientation of the magnetic moments. Therefore, the crystal
structure in the magnetic phase must become
orthorhombic.\cite{yildrim} There is, however, another role of the
orthorhombic distortion: It lifts the magnetic frustration. The
magnetic structure is stabilized by a large next-nearest neighbor
Fe-Fe interaction (along the diagonals of the Fe square lattice,
see Fig.~\ref{fig:structure}(b)), but within the tetragonal
symmetry the nearest-neighbor interactions are fully frustrated
yielding two antiferromagnetic sublattices which are completely
decoupled. Depending on the arbitrary choice of the coupling of
the two subsystems, the ferromagnetic stripes run either along the
$a$ or along the $b$ directions in orthorhombic notation. It is
thus the role of the tetragonal-to-orthorhombic phase transition
to lift the magnetic degeneracy and the frustration very similar
to the magneto-elastic coupling recently studied in
VOCl.\cite{komarek2009} The occurrence of magnetism in LaOFeAs can
be considered as an electronic nematic phase, but the magnetic
in-plane anisotropy is only a natural consequence of the fact that
the magnetic structure breaks the tetragonal rotation axis. The
similarity with liquid-crystal phases is furthermore limited, as
the high-temperature phase in LaOFeAs only exhibibits a four-fold
and not a continuous rotations symmetry. Nevertheless, one may
associate the orthorhombic phase and its precursors with an
electronic nematic state, and indeed resistivity measurements on
detwinned Ba(Fe/Co)$_2$As$_2$ crystals reveal a pronounced
electronic anisotropy.\cite{nematic}

\begin{table*}[htbp]
\caption{\label{tab:structure}Refined lattice constants, atomic
parameters and temperature factors of the LaO$_{1-x}$F$_x$FeAs
($x$=0 and 0.15) nuclear structure investigation at selected
temperatures. The respective Wyckoff sites are La $2c$
$\nicefrac{1}{4}$ $\nicefrac{1}{4}$ $z$, O/F $2a$
$\nicefrac{3}{4}$ $\nicefrac{1}{4}$ $0$, Fe $2b$ $\nicefrac{3}{4}$
$\nicefrac{1}{4}$ $\nicefrac{1}{2}$, As $2c$ $\nicefrac{1}{4}$
$\nicefrac{1}{4}$ $z$ within the tetragonal space group {\it
P4/nmm} (origin choice 2) and La $4g$ $0$ $\nicefrac{1}{4}$ $z$,
O/F $4a$ $\nicefrac{1}{4}$ $0$ $0$, Fe $4b$ $\nicefrac{1}{4}$ $0$
$\nicefrac{1}{2}$, As $4g$ $0$ $\nicefrac{1}{4}$ $z$ within the
orthorhombic space group {\it Cmme}.}
\begin{ruledtabular}
\begin{tabular}{cccccccc}
{\bf x=0} & 2 K & 100 K & 150 K & 165 K & 180 K & 200 K & RT\\
\hline $a$ (\AA) & 5.7063(4) & 5.7057(4) & 5.7000(4) & 5.6974(2) & 4.02758(4) & 4.02814(4) & 4.0322(2)\\
$b$ (\AA) & 5.6788(4) & 5.6799(4) & 5.6890(4) & 5.6920(2) & 4.02748(4) & 4.02814(4) & 4.0322(2)\\
$c$ (\AA) & 8.7094(6) & 8.7113(6) & 8.7151(6) & 8.7167(1) & 8.7191(1) & 8.7218(1) & 8.7364(4)\\
La $z$& 0.1420(4) & 0.1422(4) & 0.1428(4) & 0.1426(4) & 0.1422(4) & 0.1421(4) & 0.1416(4)\\
As $z$ & 0.6505(5) & 0.6503(5) & 0.6491(6) & 0.6493(6) & 0.6502(6) & 0.6504(5) & 0.6508(5)\\
La $B_{iso}$ & 0.07(6) & 0.17(6) & 0.18(6) & 0.24(6) & 0.26(6) & 0.25(6) & 0.65(6)\\
As $B_{iso}$ & 0.12(7) & 0.16(7)& 0.26(8) & 0.31(8) & 0.27(8)  & 0.36(8) & 0.62(8)\\
Fe $B_{iso}$ & 0.16(6) & 0.19(6)& 0.22(6) & 0.29(6) & 0.30(6)  & 0.36(6) & 0.70(5)\\
O $B_{iso}$ & 0.20(8) & 0.29(8)& 0.34(8) & 0.34(8) & 0.33(8)  & 0.42(8) & 0.84(8)\\
\hline
{\bf x=0.15}  \\
\hline $a$ (\AA) & 4.01739(3) & 4.0182(2) & - & - & - & 4.02065(3) & 4.02447(3)\\
$b$ (\AA) & 4.01739(3) & 4.0182(2) & - & - & - & 4.02065(3) & 4.02447(3)\\
$c$ (\AA) & 8.6610(1) & 8.6651(4) & - & - & - & 8.6769(1) & 8.6948(1)\\
La $z$& 0.1450(3) & 0.1450(3) & - & - & - & 0.1451(4) & 0.1450(4)\\
As $z$ &  0.6536(4) & 0.6534(4)  & - & - & - &  0.6540(4) & 0.6540(4) \\
La $B_{iso}$ & 0.37(5) & 0.36(5) & - & - & - & 0.54(6) & 0.73(5)\\
As $B_{iso}$ & 0.18(6) & 0.26(6) & - & - & - & 0.40(7) & 0.58(6)\\
Fe $B_{iso}$ & 0.27(4) & 0.27(4)& - & - & -  & 0.44(5) & 0.66(4)\\
O/F $B_{iso}$ & 0.42(6) & 0.35(6)& - & - & -  & 0.41(7) & 0.73(7)\\
\end{tabular}
\end{ruledtabular}
\end{table*}

\begin{figure}
\includegraphics[width=0.45\textwidth]{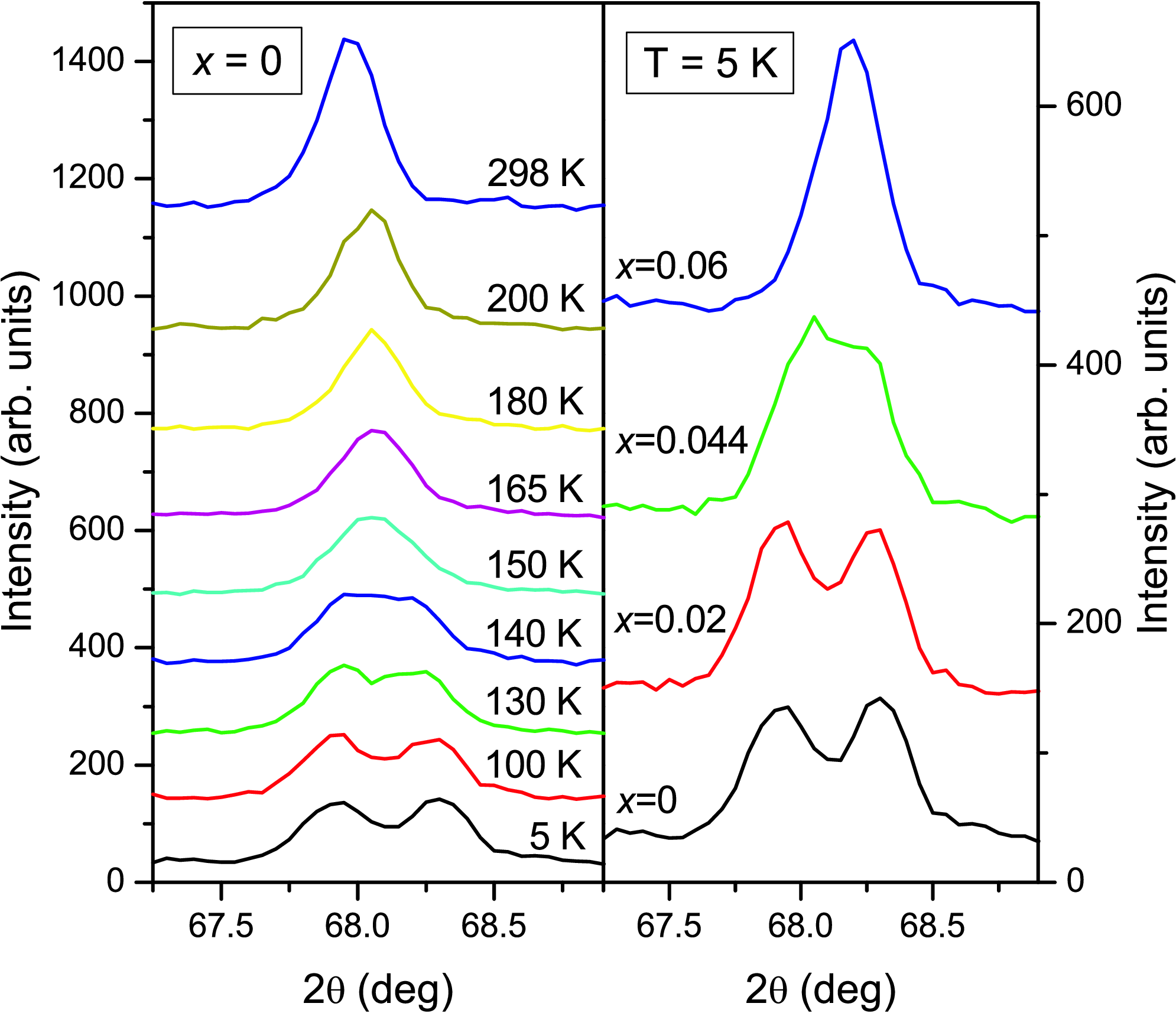}
\caption{\label{fig:220_T} (Color online) The structural phase
transition is revealed by the temperature (left panel) and doping
dependent (right panel) splitting of the tetragonal (220)
reflection into the orthorhombic (400) and (040) reflections.}
\end{figure}

At low temperature the orthorhombic splitting in LaOFeAs is well
documented for example by the neutron diffraction profiles of the
(400)/(040) reflections, see Figure~\ref{fig:220_T}. Close to the
structural transition these two reflections, however, overlap due
to the smaller orthorhombicity in the neutron and in the X-ray
experiments. The neutron data were used to refine the orthorhombic
structure model up to 180\ K and with the X-ray data we fitted the
total width of the (400)/(040) scattering. Considering the
steepest temperature dependencies of the orthorhombicity and of
the peak width, one may determine the structural transition
temperature to about $T_s$$\sim$150\ K, which perfectly coincides
with the well-defined kink in the resistivity \cite{kon2009} and
with the maximum in the thermal expansion
coefficient,\cite{wang2009} both measured on nearly identical
samples. However, the broadening of the (400)/(040) scattering as
well as a finite orthorhombicity remain clearly visible above
$T_s$, they only vanish at a temperature of about 200\ K. These
findings fully agree with similar diffraction studies by McGuire
et al. \cite{mcg2008} and by Nomura et al. \cite{nomura2009}
indicating that this feature is not sample specific but an
intrinsic property of LaOFeAs. Great care is thus needed to
determine the {\it true} structural phase transition temperature
in LaO$_{1-x}$F$_x$FeAs with the aid of diffraction data only. The
broadening of the peaks in the intermediate temperature range
should be interpreted as an inhomogeneous phase with strong local
orthorhombic distortions as precursors of the long-range
transition. The correlation length of the local orthorhombic
distortions must be rather large, of the order of the coherence
length of the radiation, i.e. of the order of several hundreds of
Angstroems. The broad temperature range, where orthorhombic
precursors exist, seems to coincide with the upturn of the thermal
expansion coefficient well above the long-range structural
transition \cite{wang2009} and with the upturn
in the electric resistivity.\cite{kon2009} 
Evidence for orthorhombic precursors can also be deduced from the
resistivity measurements on detwinned Ba(Fe/Co)$_2$As$_2$
crystals, as the anisotropy sets in already above the long-range
structural transition.\cite{nematic}

\begin{figure}
\includegraphics[width=0.45\textwidth]{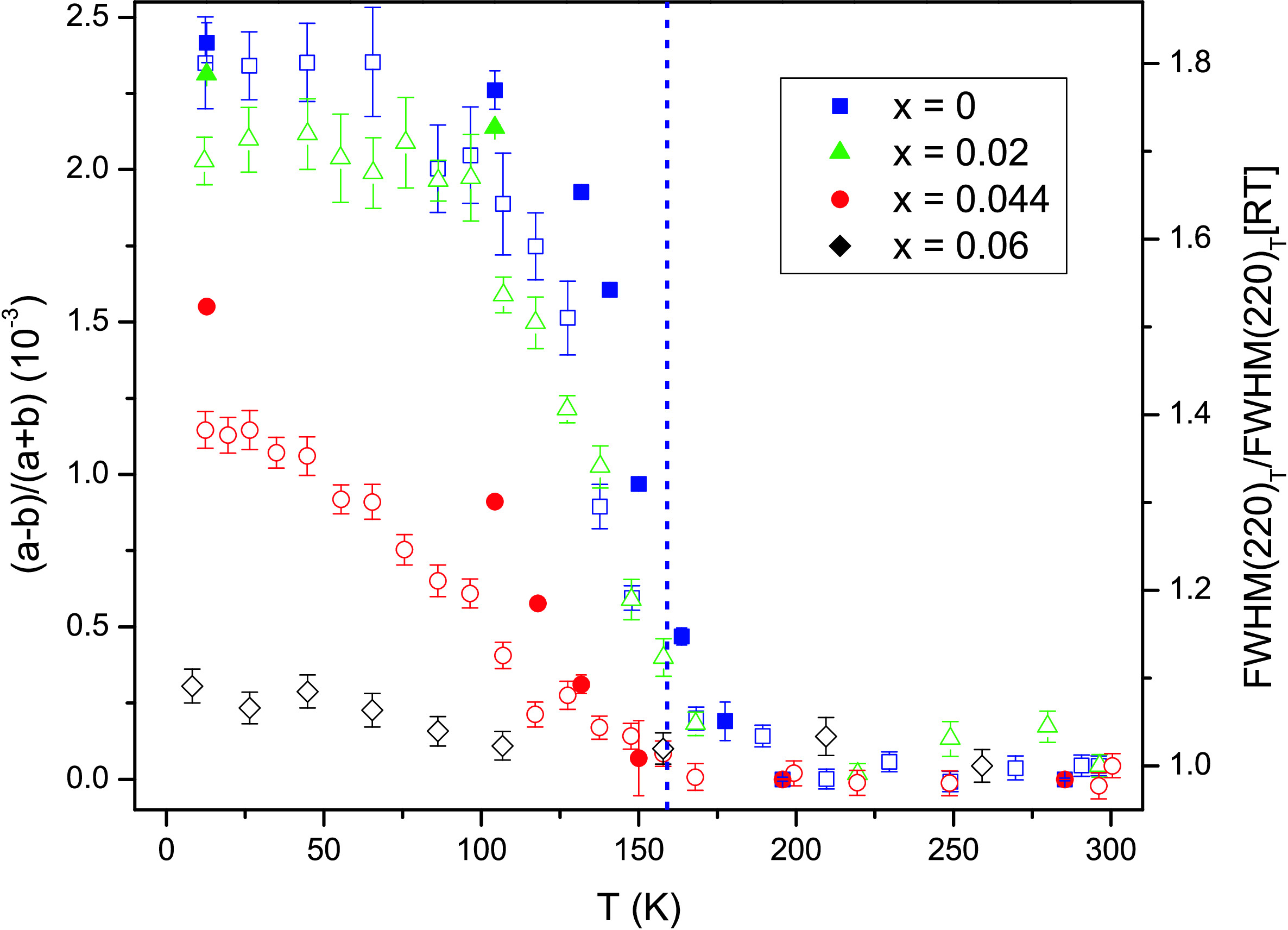}
\caption{\label{fig:ortho} (Color online) The tetragonal to
orthorhombic phase transition expressed by the orthorhombicity
$(a-b)/(a+b)$ (filled symbols, neutron diffraction) and the FWHM
of the splitting (220)$_T$ reflection (open symbols, X-ray
diffraction); the vertical dashed line indicates the structural
transition temperature indicated by the resistivity and thermal
expansion anomalies in LaOFeAs.}
\end{figure}

The high-resolution neutron data were used to refine structure
models varying the cell constants, the $z$ component of the La and
As positions and the isotropic temperature factors of all species.
From the refined parameters (see Tab.~\ref{tab:structure}) further
structural aspects like interatomic distances and angles have been
deduced, see Fig.~\ref{fig:param}.

The temperature dependence of several structural aspects is shown
in Figs.~\ref{fig:param}(c) and (d) for $x$=0 and $x$=0.15,
respectively. The lattice constants show an expected dependence on
temperature, but the remaining parameters reveal interesting
features around the structural phase transition. In LaOFeAs, the
interatomic distances and especially the tetrahedra angles reveal
a strong discontinuity between 140 K and 180 K, which can be
correlated to the onset of magnetic order (see
Fig.~\ref{fig:moment}) and the occurrence of the orthorhombic
precursors (see Fig.~\ref{fig:ortho}), respectively.

As this discontinuity might be a crucial result it has been
assured that it is not a result of artifacts emerging from the
refinement using the orthogonal unit cell instead of the
tetragonal one. Therefore, all diffraction patterns have been
analyzed as well with the microstrain option implemented in
FullProf, which introduces an orthorhombic distortion into a
tetragonal cell. The parameters obtained by this refinement method
proved to be equivalent within the error bars. Additionally, a
diffraction pattern from within the transition regime has been
analyzed using two refined structural models of a tetragonal (just
before the transition) and an orthogonal phase (just after the
transition) by simply refining the respective scale factors. This
procedure led to a considerably worse agreement implying that the
broadening of the structural transition is a system-inherent
property and not due to inhomogeneities in the chemical
composition.

The orthorhombic precursors and the structural anomalies appear to
be closely related with the electronic and magnetic properties. In
the temperature range of the orthorhombic precursors the
resistivity shows an upturn, whereas the long-range orthorhombic
transition finally induces a decrease in resistivity, see for
example Ref.~\onlinecite{kon2009}. The magnetic importance of the
structural transition is visible in the reduction of the magnetic
susceptibility which can be interpreted by either a suppression of
ferromagnetic fluctuations or an increase of the antiferromagnetic
interaction. It appears interesting to note that the structural
anomalies observed concern those structural parameters which are
determining the size of the Fe moment, i.e. the FeAs distance and
the FeAs layer thickness.\cite{yin2008} Furthermore, the
structural anomalies indicate extrema in the temperature
dependencies of the respective parameters, whereas a simple
structural transition should induce a kink or a jump in the
temperature dependency. These extrema, however, agree nicely with
the anomalous contribution to the thermal expansion, which
exhibits a sign change in the same temperature range and thus also
yields an extremum for the anomalous component of the lattice
volume.\cite{wang2009}

\subsection{ Doping dependence of the crystal structure in LaO$_{1-x}$F$_x$FeAs }

The tetragonal to orthorhombic transition has been studied by
neutron diffraction in detail for two samples with $x$=0 and 0.045
and is represented in Fig.~\ref{fig:ortho} by the orthorhombicity
$(a-b)/(a+b)$. From the additional X-ray diffraction experiments,
also including the $x$=0.02 and 0.06 samples, the temperature
dependent evolution of the FWHM of the splitting (220)$_T$
reflection has been extracted, where the obtained values have been
normalized to the FWHM at RT.

It can be seen for $x\leq 0.045$ that the splitting of the {\it a}
and {\it b} lattice constants in an orthorhombic fit, and
therefore the broadening of the affected Bragg peaks, sets in well
above the transition temperatures given in the
literature.\cite{nom2008,mcg2008,lue2009} No high-resolution
neutron data  has been collected for the $x$=0.02 sample between
100 K and 200 K, but the increase of the FWHM obtained by the
X-ray experiments yields a broad transition regime very similar to
the observation in the pure compound. The smearing of the
structural phase transition by orthorhombic precursors is thus
present over a finite doping interval. At the intermediate doping
$x$=0.045, it is difficult to determine the structural transition
temperature with the diffraction data, as the transition between
the precursors and the long-range orthorhombic phase seems to be
quite sluggish. The $x=0.06$ sample does not exhibit a comparable
peak broadening indicating that the orthorhombic distortion is
suppressed in the superconducting phase in LaO$_{1-x}$F$_x$FeAs.
Note that also the previously studied superconducting sample with
$x=0.05$ did not show the large peak broadening \cite{lue2009} in
contrast with the non-superconducting $x=0.045$ sample studied
here. In spite of the strong precursors associated with the
structural transition, the long-range orthorhombic distortion
seems to become very rapidly suppressed near $x=0.05$ i.e. at the
boundary between superconducting and non-superconducting samples.
However we may not fully rule out that some orthorhombic
precursors persist into the superconducting phase.

\begin{figure*}
\includegraphics[width=0.24\textwidth]{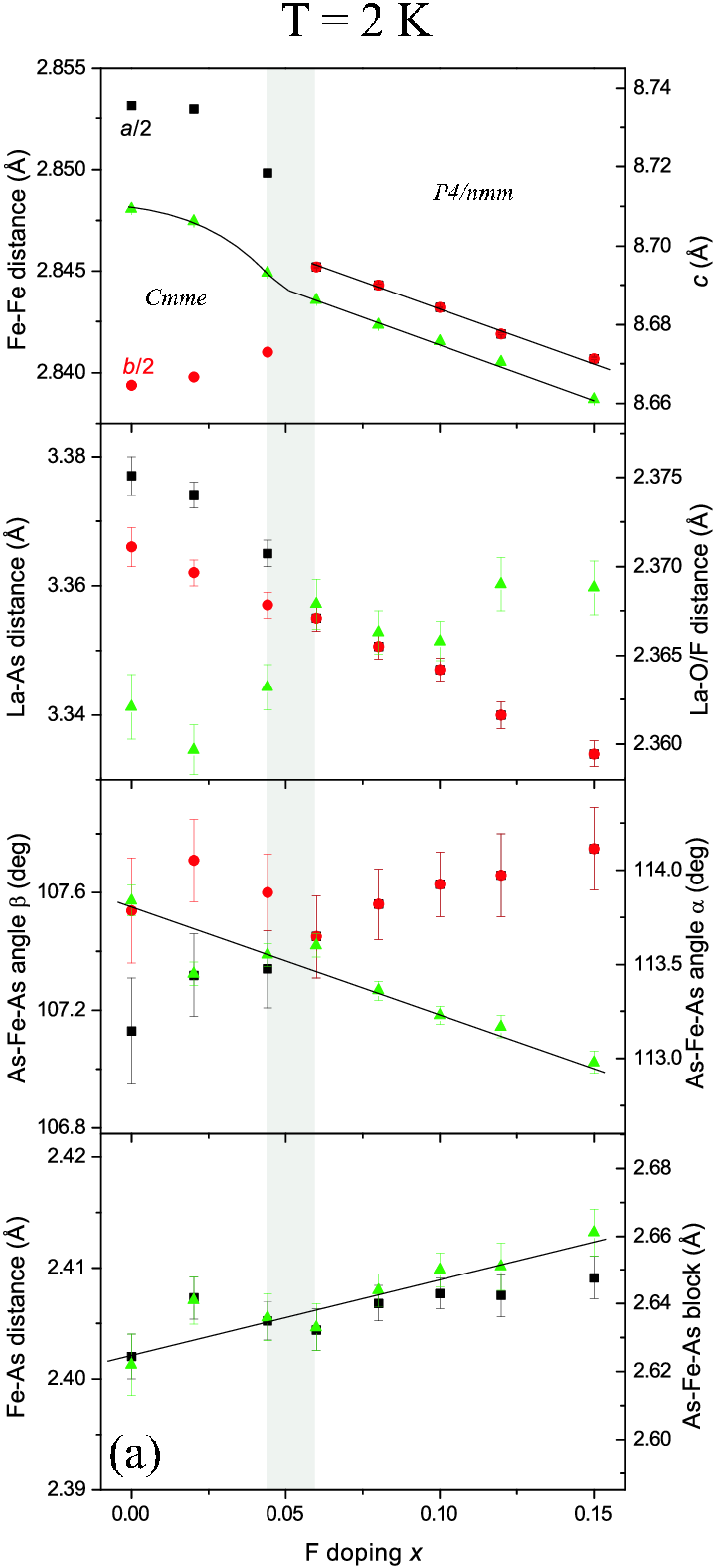}
\includegraphics[width=0.24\textwidth]{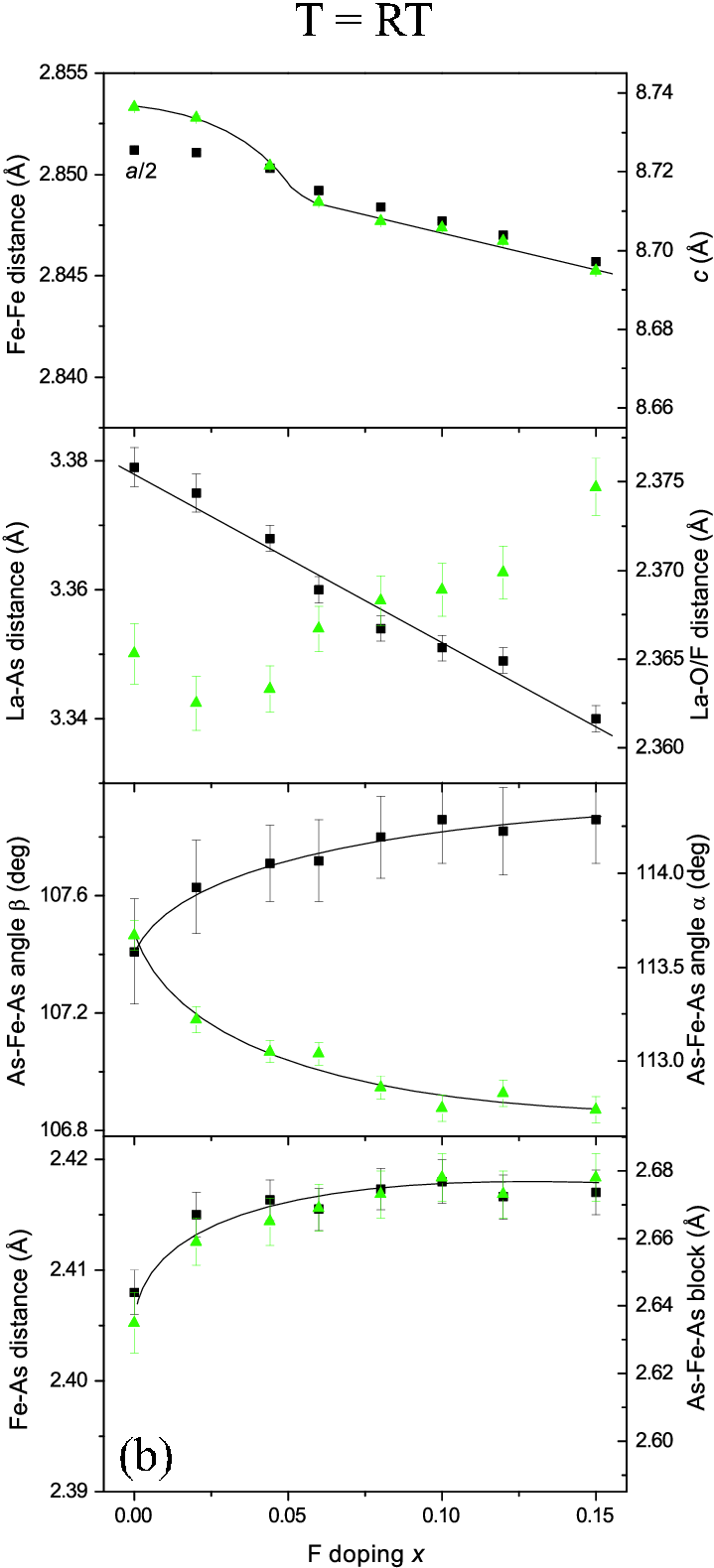}
\includegraphics[width=0.24\textwidth]{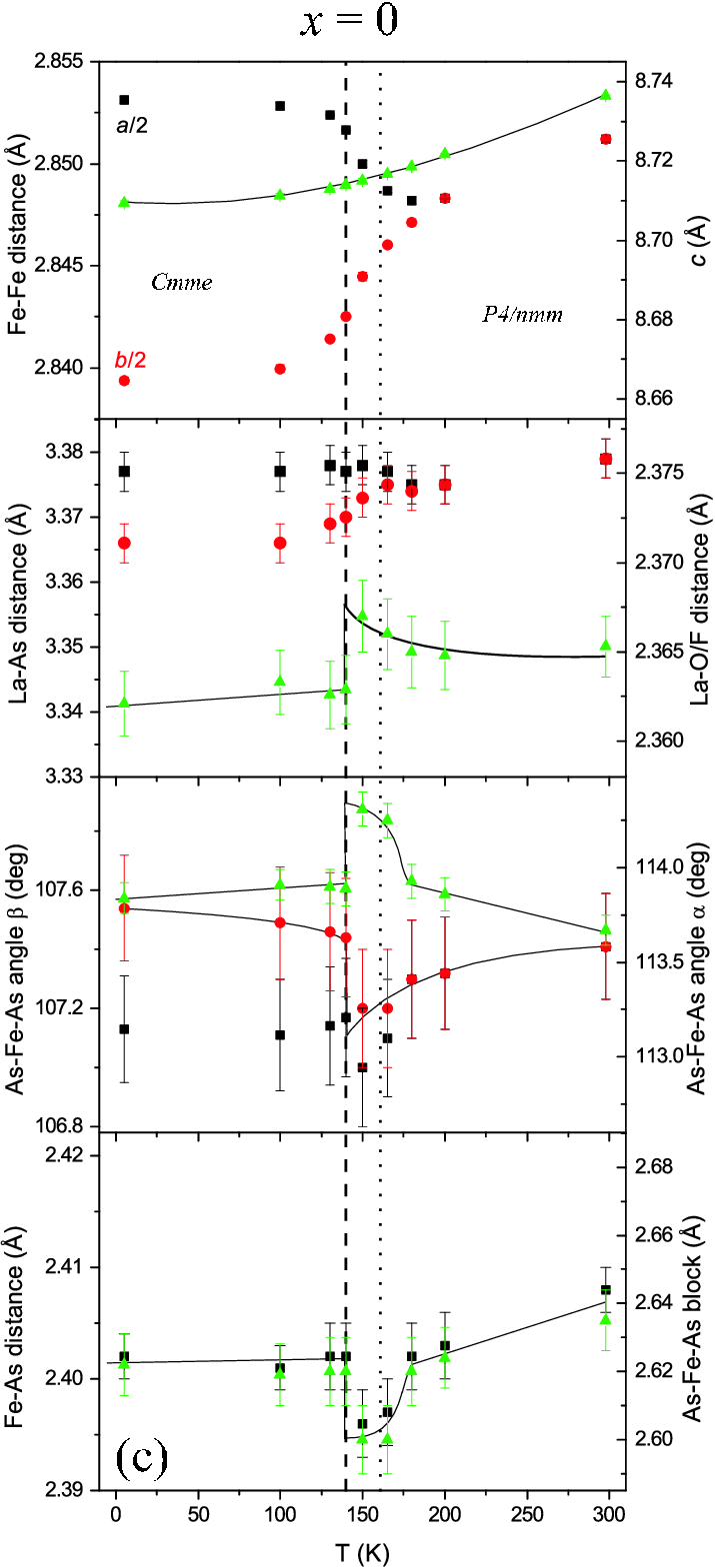}
\includegraphics[width=0.24\textwidth]{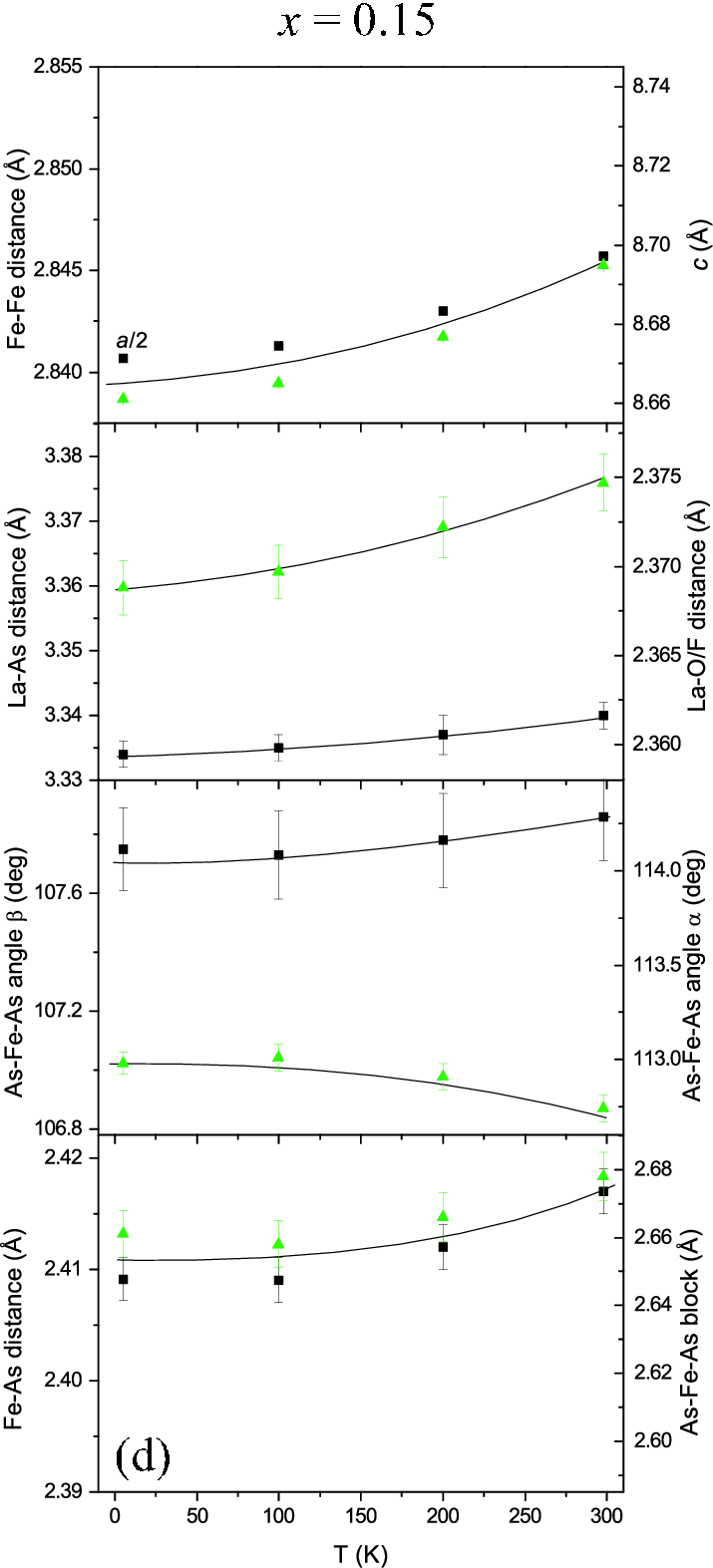}
\caption{\label{fig:param} (Color online) Structural parameters of
the LaO$_{1-x}$F$_x$FeAs compounds as a function of F doping [(a)
T=2 K, (b) T= RT] and temperature [(c) $x$=0, (d) $x$=0.15].
(Black) squares and (red) dots correspond to the left abscissa,
while (green) triangles belong to the right one. The grey shaded
area in (a) denotes the doping dependent structural phase
transition regime. In (c) the dashed line marks the onset of
static magnetism while the dotted line indicates the structural
phase transition as taken from Ref.~\onlinecite{kon2009}. Solid
lines are guides to the eye.}
\end{figure*}

The structural phase transition as a function of F doping is
located at $0.045<x<0.06$, evidenced by the merging of the
different Fe-Fe/La-As distances and As-Fe-As angles, which is
shown in Fig.~\ref{fig:param}(a) for T=2 K. The doping dependent
evolution of the structural parameters does not exhibit any
significant differences between T=2 K and RT
(Fig.~\ref{fig:param}(b)), although the structural phase
transition is only observed below 150 K. The lattice parameter
{\it c} as well as the La-As distance decrease monotonically as a
consequence of the charge carrier injection causing a Coulomb
attraction of the LaO/F and FeAs layers. On the other hand the
charge carrier injection into the FeAs layer leads to an increase
of the Fe-As distance and of the As-Fe-As block thickness. The
doping causes the FeAs$_4$ tetrahedra to become more homogeneous
which is expressed by the different As-Fe-As angles approaching
the perfect angle of 109.47$^\circ$. Our findings are similar to
those presented in Ref.~\onlinecite{hua2008} but extending to
higher F concentrations.

The temperature dependence of the same structural aspects is shown
in Figs.~\ref{fig:param}(d) for $x$=0.15. For this superconducting
sample we find no structural anomaly in the temperature
dependencies.

\section{Conclusions}

We have performed high-resolution and high-flux neutron powder
diffraction experiments on the oxypnictide series
LaO$_{1-x}$F$_x$FeAs, which on the one hand confirm the structural
parameters and their evolution with F-doping reported previously,
but on the other hand our results indicate a structural anomaly
and a larger magnetic order parameter.

With the high-resolution and good statistics of the diffraction
data taken on the D20 diffractometer, we may unambiguously
determine the magnetic structure in LaOFeAs obtaining a Fe
magnetic moment of 0.63(1)~$\mu$\textsubscript{B} which is in good
agreement with a previous NMR study,\cite{grafe2009} but about a
factor two higher than previous neutron diffraction, muon spin
relaxation and M\"o\ss bauer reports. The Fe magnetic moment has
been subject of debate as it is largely reduced compared to the
theoretically expected value of almost
2~$\mu$\textsubscript{B}.\cite{yin2008} Furthermore, the moment
alignment is along the {\it a} axis parallel to the stacking
direction of the ferromagnetic Fe stripes, which fully agrees with
the magnetic structure observed in other FeAs compounds. The 2\%
doped sample exhibits a similarly large magnetic moment, but from
the magnetic reflection profile, i.e. from a broader FWHM, a
perturbation of the static magnetic order can be deduced. We still
find a sizeable ordered moment for a non-superconducting sample
with $x=0.045$ in good agreement with muon spin relaxation and
M\"o\ss bauer results.\cite{lue2009} Since this doping level is
very close to the superconducting part of the phase diagram, one
may deduce a well-defined phase boundary in between.

Concerning the structural properties, we have observed that the
tetragonal-to-orthorhombic phase transition extends to a rather
large temperature regime with pronounced precursors persisting
well above the long-range transition temperature. It is difficult
to determine the transition temperature by diffraction techniques
as long-range and short-range distortions are nearly impossible to
separate. The analysis of the peak height of the nuclear Bragg
peaks is certainly misleading. Similar precursor effects should
also exist for the ReO$_{1-x}$F$_x$FeAs materials and might be one
reason for discrepancies in the phase diagrams obtained by
different groups. In LaO$_{1-x}$F$_x$FeAs the long-range
orthorhombic distortion is fully suppressed by amounts of doping
which are below the level needed to induce superconductivity.
Taking further account of the magnetic neutron diffraction
results, we confirm the phase diagram presented in
Ref.~\onlinecite{lue2009}. However, we may not exclude that some
orthorhombic precursors persist into the superconducting phase in
LaO$_{1-x}$F$_x$FeAs as well as in other ReO$_{1-x}$F$_x$FeAs
series.\cite{zha2008}

In pure and low-doped LaO$_{1-x}$F$_x$FeAs, the onset of static
magnetism as obtained by muon spin relaxation\cite{lue2009} and by
the temperature dependence of the magnetic Bragg peaks lies within
the broad regime of the tetragonal-to-orthorhombic transition. For
pure LaOFeAs, we find structural anomalies just above the onset of
magnetism, as the FeAs distance and the FeAs layer thickness pass
through a minimum in agreement with the effect observed for the
thermal-expansion coefficient. These features seem to arise from
the strong magnetoelastic coupling between the shape of the FeAs
tetrahedra and the size of the magnetic moment suggesting that the
enhancement of the antiferromagnetic correlations through the
orthorhombic distortion implies a variation of the magnetic
moment.

\begin{acknowledgments}
This work was supported by the Deutsche Forschungsgemeinschaft
through the Sonderforschungsbereich 608.
\end{acknowledgments}


\begin{thebibliography}{31}
\bibitem[{\citenamefont{}(2010)\citenamefont{}}]{note}
The propagation vector $\mathbf k$=(0 1 $\frac{1}{2}$) is an
equivalent option.
\end{thebibliography}

\end{document}